\documentclass[acmsmall]{acmart} 
\usepackage[utf8]{inputenc}

\setcopyright{cc}
\setcctype{by}
\acmJournal{PACMHCI}
\acmYear{2026} \acmVolume{10} \acmNumber{7} \acmArticle{GAMES076}
\acmMonth{11} \acmDOI{10.1145/3831353}

\acmConference[CHI PLAY '26]{The Annual Symposium on Computer-Human Interaction in Play}{November 02--05, 2026}{York, United Kingdom}

\acmISBN{}

\AtBeginDocument{%
  }
    
\begin{document}

\title{Understanding Game Coaching on Gig Platforms}


\author{Hwijoon Lee}
\email{lee.hw@northeastern.edu}
\affiliation{%
  \institution{Northeastern University}
  \city{Boston}
  \state{Massachusetts}
  \country{USA}}
  
\author{Saiph Savage}
\email{s.savage@northeastern.edu}
\affiliation{%
  \institution{Northeastern University}
  \city{Boston}
  \state{Massachusetts}
  \country{USA}
}

\renewcommand{\shortauthors}{Lee et al.}

\begin{abstract}
Freelance game coaches monetize their gaming expertise by offering personalized instruction to players seeking to improve, working through gig platforms, yet little is known about how they operate. To address this gap, we conducted semi-structured interviews with 20 experienced freelance coaches across 17 competitive games on Fiverr. Despite lacking shared formal training, these coaches converged on similar practices centered on rapport-building, individualized diagnosis, and adaptive feedback. We identify two structural conditions shaping this work: dual precarity, in which coaches navigate both gig platform instability and the lifecycle volatility of live-service games; and earned authority, in which coaches must continually establish legitimacy through visible competitive achievement within the same gaming spaces as their students. These coaches welcomed AI for administrative and analytic support but resisted its use in live interactions where trust, relational engagement, and situated judgment remained central. We discuss implications for Games HCI and the design of computational coaching systems.
\end{abstract}




\received{February 2026}
\received[revised]{June 2026}
\received[accepted]{July 2026}

\maketitle

\section{Introduction}
Competitive multiplayer games depend on broader ecosystems of learning, mentorship, and expertise development that extend well beyond gameplay itself \cite{kow2013media, kow2014crafting, kleinman2024backseat}. Within these ecosystems, game coaching, in which more skilled players teach others to improve their play, has become a central way that players develop competitive expertise. Yet research on game coaching has concentrated on professional team settings \cite{watson2025introducing, lee2025crafting}, leaving everyday, non-institutional forms of coaching largely unstudied. Understanding how coaching is practiced in these everyday settings is a prerequisite for designing tools that fit how competitive games are taught and learned.

Computational and AI-powered tools are increasingly entering competitive gaming ecosystems, promising automated gameplay analysis, personalized training recommendations, performance tracking, and real-time feedback \cite{kleinman2022kills, wang2024feature, wallner2021players}. Yet recent work shows that integrating these tools into actual coaching practice is far from straightforward, with players and coaches reporting skepticism, difficulty interpreting outputs, and limited engagement \cite{kleinman2026really}. Building on these prior works, we see an opportunity to examine how coaching operates outside institutional settings, since a clearer account of everyday coaching practice would help clarify how these tools should support players.

Freelance game coaching services are increasingly appearing on gig platforms, where experts monetize their competitive skills by offering personalized lessons to players seeking to improve their rank, game knowledge, and performance. These services operate through algorithmically managed digital labor platforms \cite{wood2019good} rather than within traditional esports organizations \cite{watson2025introducing}. To understand this practice where it has stabilized, we focus on experienced coaches, who have had time to develop consistent working methods and accumulate knowledge of what helps players improve. Examining how they structure their work, the challenges they face, and the role technology plays in it is a necessary step toward tools that fit freelance coaching.

To understand freelance game coaching practice, we conducted semi-structured interviews with 20 experienced coaches working across 17 competitive games on a gig platform (Fiverr). Focusing on coaches who have built a sustained presence on the platform, we examine how they structure sessions, diagnose player weaknesses, manage student relationships, and navigate platform constraints. Our findings contribute to Games HCI by showing how coaching practices are shaped by the games themselves, their mechanics, metagame cycles, and shifting player needs. We further identify two structural conditions that sustain freelance coaching, dual precarity and earned authority, and use them to characterize the human labor, pedagogy, and platform conditions underlying game coaching ecosystems.

We uncovered that although coaches lacked shared formal training, they independently converged on a recognizable pedagogical practice centered on rapport-building, individualized diagnosis, and adaptive feedback. Rather than primarily repeating information already available through online guides or videos, coaches created value through situated interpretation: understanding a player’s gameplay, habits, emotional state, and learning needs within the specific context of play. These findings extend Games HCI research on learning, expertise, and player support by showing how video game coaching operates as an interactive, relational, and context-dependent form of gameplay assistance \cite{kleinman2022kills,kleinman2024backseat,lee2025crafting}. 

Our analysis also surfaces two structural conditions shaping game coaching. The first is ``dual precarity'': coaches simultaneously navigate gig platform algorithmic instability and the lifecycle volatility of specific game titles, requiring continuous adaptation to platform visibility systems, shifting optimal strategies known as the metagame \cite{kokkinakis2021metagaming}, and fluctuating player populations \cite{demediuk2018player}. This highlights how coaching is tightly coupled with gig platform infrastructures and evolving game ecologies. The second is ``earned authority'': unlike institutional educators who rely on formal credentials, freelance coaches must continuously establish legitimacy through visible competitive achievement within the same gaming spaces their students inhabit. Credibility therefore emerges through ongoing performance, interaction, and demonstrated expertise within play communities themselves. These conditions also shaped coaches’ perspectives on AI-supported tools. Coaches welcomed AI systems that supported administrative and preparatory labor, such as scheduling, communication, and training plan generation, but resisted AI replacing live coaching or gameplay diagnosis, which they viewed as requiring situated judgment and interpersonal trust.

Through these findings, we contribute to Games HCI in three ways. First, we introduce dual precarity as a structural condition that characterizes freelance game coaching within platform-mediated gaming economies. Second, we develop the concept of earned authority to explain how instructional legitimacy is established and maintained in competitive gaming environments. Third, we articulate implications for the design of computational coaching systems, showing that while AI and other computational tools may productively augment coordination and preparation work, the relational and diagnostic core of game coaching remains deeply tied to situated human expertise, social interaction, and the lived context of gameplay.
\section{Related Work}

\subsection{Platform Labor and Reputation}
Digital labor platforms restructure work by organizing it through marketplace-based matching systems rather than traditional employment relationships \cite{vallas2020platforms, kenney2016rise, carlos2021lose}. Prior research on remote gig work characterizes this arrangement as structurally precarious: income is volatile, contracts are short-term, and institutional protections are limited \cite{wood2019good}. Beyond economic instability, gig workers face challenges of professional legitimacy: in the absence of organizational affiliation, credibility is not institutionally conferred but must be actively constructed and continuously maintained by the worker \cite{gandini2016digital}.

This restructuring is operationalized through systems of algorithmic management \cite{lee2015working, vallas2020platforms}. Platforms do not merely facilitate transactions; they actively govern visibility and opportunity by ranking profiles, sorting search results, and quantifying performance through ratings and behavioral metrics \cite{lee2015working}. As Wood describes, workers experience a form of “algorithmic control” in which platform-generated metrics shape search placement and future job access \cite{wood2019good}. Visibility and continued access to work are therefore shaped by ongoing platform evaluation rather than secured through stable employment status \cite{wood2019good}.

Within this configuration, workers must actively manage how they are perceived. Gandini \cite{gandini2016digital} argues that digital labor markets transform freelancers into ongoing self-branding projects, requiring them to curate profiles, accumulate positive reviews, and strategically present their skills to remain competitive \cite{gandini2016reputation}. Because visibility and opportunity are algorithmically distributed, reputation becomes a resource that must be continuously cultivated \cite{gandini2016reputation}. In the absence of stable institutional credentials, credibility is constructed through platform-mediated signals and sustained through persistent reputational labor \cite{gandini2016digital}. 

\subsection{Gigified Knowledge Services}
Gig economy platforms have expanded beyond ride-sharing and microwork to encompass knowledge-based professions \cite{carlos2021lose, kassi2018online}. In these contexts, the reputational dynamics and platform constraints described above become especially consequential, as service quality is more difficult to assess and provider–client relationships play a larger role in determining outcomes \cite{gandini2016reputation, carlos2021lose}.

Unlike microwork, which typically involves standardized tasks requiring no specialized expertise and minimal collaboration between worker and requester \cite{margaryan2019workplace, gray2019ghost}, knowledge-based gig work demands domain-specific skill and sustained provider–client collaboration to complete complex, individually scoped projects \cite{carlos2021lose}. Because the outcomes of such services are difficult to evaluate in advance, clients rely heavily on platform-mediated signals such as ratings and reviews when selecting providers \cite{gandini2016reputation}. Yet these signals capture transactional satisfaction rather than the relational and diagnostic dimensions of quality that distinguish knowledge services from routine task completion \cite{carlos2021lose, jarrahi2020platformic}.

Teaching and tutoring have also emerged as gig-mediated knowledge services within platform economies \cite{carlos2021lose, kassi2018online}. In these domains, expertise is both domain-specific and relationally enacted, requiring sustained interaction between provider and client to diagnose needs and co-produce outcomes \cite{carlos2021lose, jarrahi2020platformic}. Xia et al. found that frequent tutor rotation on language platforms was associated with both reduced relational continuity and slower improvements in speaking performance \cite{xia2022understanding}. Curran showed that learners on gig language platforms emphasized interpersonal warmth and conversational rapport when describing their ideal teachers, alongside persistent ideologies of native speakerism \cite{curran2023more}. Zhang et al. documented how transnational gig-education platforms required teachers to perform emotional labor across cultural contexts while simultaneously navigating platform-imposed constraints such as algorithmic evaluation and surveillance \cite{zhang2025identity}.

Collectively, this literature demonstrates that gig-mediated educational services are shaped by platform governance, reputational infrastructures, and relational labor. In parallel, AI has already been examined as a means of supporting personalized instruction: work on intelligent tutoring systems shows that AI can deliver individualized diagnosis and feedback at scale \cite{vanlehn2011relative}, while research on human–AI augmentation has examined how AI tools support rather than replace human instructors \cite{holstein2018student}. Yet this work, too, has focused on traditional educational settings, leaving its applicability to gig-mediated game coaching services unexamined.

As esports has evolved into a global and increasingly structured competitive ecosystem \cite{scholz2020deciphering}, coaching has expanded beyond professional team settings and become visible on digital marketplaces. For example, as of 2026, over 2,500 active listings related to game coaching are publicly visible on the freelance platform Fiverr, indicating the presence of coaching as a form of platform-mediated gig work. Despite this growing visibility, academic research examining gig-based game coaching practices remains limited.

\subsection{Competitive Game Ecosystems and Coaching}
Previous research has examined how activities within online games can be organized as income-generating labor. Early studies of gold farming established that the production of in-game currencies, items, and services constitutes a recognizable form of game-based labor rather than incidental extensions of play \cite{heeks2009understanding}. More recent scholarship extends this line of work to platform-mediated game services. Drawing on an ecosystem-level account of game labor, Johnson and Woodcock demonstrate how the rapid expansion of competitive gaming has generated new forms of work while simultaneously producing precarious conditions across the actors who sustain the scene \cite{johnson2021work}. Together, these studies situate game-related services within broader dynamics of platform labor, underscoring how the commodification of gameplay reshapes labor processes, credibility, and value creation in game-based marketplaces.

Within this broader landscape, coaching has attracted growing attention as a knowledge-intensive practice within competitive game ecosystems. Watson et al. interviewed professional head coaches to identify structural conditions shaping their work, including constant tactical change driven by frequent game patches and pre-match drafts that force coaches to revise strategy under win-loss pressure from team owners and global audiences. A further finding is the lack of formal training or career paths for esports coaches, who learn to coach through their own playing experience, informal mentoring, and community resources. This absence tends to narrow their teaching toward direct instruction, providing immediate solutions rather than developing players' independent capacity to diagnose and adapt \cite{watson2025introducing}. Lee et al. bring an HCI lens to this domain, conducting an observational study at an elite South Korean training academy that surfaces two recurring difficulties in the coaching process. The first is managing the volume of information coaches must track when overseeing multiple players over extended periods; the second is the limited toolkit self-taught coaches bring to motivating students and managing their emotions \cite{lee2025crafting}.

Alongside research on coaches, games HCI has examined how computational and AI-driven tools can support player learning directly. Kleinman et al. mapped existing computational esports tools onto phases of self-regulated learning and identified open opportunities for computational support across before, during, and after game play \cite{kleinman2022time}. Wang et al.'s feature analysis of fifteen commercial live-companion tools for League of Legends and Valorant extends this picture: while data-driven features such as real-time statistics, post-match reflections, and playstyle summaries are now widespread, the role of AI in these tools remains limited and largely opaque \cite{wang2024feature}. Returning to the coaching context, Lee et al.'s observation of an elite training academy further points to AI assistants that record game state and prior coaching input as a concrete site for tool development \cite{lee2025crafting}. However, this body of work has examined AI-supported learning either for individual players or within elite team and academy settings, leaving freelance coaching practice unexamined.

Across these strands — platform labor, gigified knowledge services, game-related labor, esports coaching, and AI-supported player learning — freelance game coaching sits at an intersection that none has examined directly. Theoretical frameworks for skill acquisition and talent development in games have been proposed~\cite{bubna2023coaching}, and coaching research has begun to examine salaried roles within professional team or academy settings, but empirical accounts of how independent coaches actually practice remain absent. A growing number of coaches now operate through freelance marketplaces, navigating algorithmic visibility, delivering personalized instruction, and grounding their authority in in-game achievement rather than institutional credentials. Alongside this growth, AI tools are beginning to enter the game learning environment, extending prior computational learning support into more adaptive and individualized forms. These include platform-side AI features such as Fiverr's AI auto-reply system, and game-specific AI services such as Stockfish, which has become a routine part of coaching in chess platforms, and Osirion, an AI-powered Fortnite gameplay analysis platform that coaches in our sample reported using. How freelance coaches encounter and interpret these emerging tools therefore forms a relevant extension of prior computational learning tool research. Our work investigates these freelance coaches, offering an empirical account of game coaching as platform-mediated gig work.

\section{Methods}
To understand the practices, challenges, and perspectives of experienced game coaches working on freelance platforms, we conducted IRB approved semi-structured interviews with 20 coaches. This qualitative approach allowed us to surface insights about an understudied population and context. Our research team includes members with prior experience in competitive gaming communities and direct familiarity with gig platform ecosystems as both users and researchers. No member of the team had a prior relationship with any participant.

\subsection{Participants}
We recruited 20 experienced game coaches (19 male, 1 female; aged 19–33) who were actively offering coaching services on Fiverr, a major freelance marketplace. To capture diverse perspectives within the game coaching ecosystem, we deliberately recruited coaches across a range of competitive game titles, ultimately spanning 17 games across multiple game genres. Gender was not a recruitment criterion, and coach gender is often not visible on Fiverr listings if profiles do not include face photos or gendered identifiers. The resulting sample composition therefore reflects the available freelance game coaches on the platform rather than purposive sampling decisions. Detailed participant demographics are presented in Table~\ref{tab:demographics}. As shown in Table~\ref{tab:demographics}, our sample includes chess coaches. While chess may appear atypical alongside contemporary video game titles, it ranks among the top six game titles by active coach count in Fiverr's game coaching category \cite{fiverr2026gamecoaching}, making it an important part of the freelance game coaching ecosystem we set out to study.

Recall that this study focuses on experienced freelance game coaches because they can be well positioned to reveal how coaching works as a stable practice and what role technology can play within it. Recruiting participants who met this criterion required a sampling approach suited to that goal. We therefore followed established guidance on ``purposive sampling of information-rich cases'' \cite{patton2002two}, a strategy in which participants are deliberately selected not for statistical representativeness but because they are likely to yield deep and substantive insight into the phenomenon under study. Applied here, this meant targeting coaches with demonstrable accumulated experience rather than those who were new to coaching.
However, operationalizing experience on Fiverr presented a practical challenge, as the platform does not expose a coach's session counts or earnings. Following prior gig-work research that uses platform-based activity measures as proxies for accumulated experience \cite{zwettler2024kicking, savage2020becoming}, we required participants to have at least 30 verified client reviews. Review count was the only practice-volume signal Fiverr exposes publicly, and without such a threshold a coach with few or no reviews could have completed anywhere from zero to dozens of unreviewed coaching sessions, making it difficult to establish that participants were drawing on substantive experience when discussing their workflows, strategies, and platform navigation. The threshold helped us to ensure that interviews were grounded in accumulated coaching practice rather than initial experimentation.

We contacted eligible coaches through Fiverr's messaging system with a recruitment notice that disclosed the study's purpose, the voluntary nature of participation, and the interview format. We arranged compensation through Fiverr at each coach's listed hourly rate, booked as a paid interview rather than a coaching session. At the start of each session, we verbally reviewed the study procedures, explained that participants could decline to answer any question or withdraw at any point without penalty, and described how recordings and transcripts would be stored and used. We then obtained verbal consent before recording began. This study was reviewed and approved by the Institutional Review Board at our institution.

\begin{table*}[t]
\footnotesize
\caption{Characteristics of participants. All participants actively coached on Fiverr at the time of interview.}
\Description[Demographics of 20 game coaches]{Table listing 20 participants with their gender, age, nationality, game coached, student count, years of experience, education level, and coaching status. 19 male and 1 female, aged 19 to 33, from 16 countries, coaching 17 different games.}
\label{tab:demographics}
\begin{tabular}{l l c l l r c l l}
\hline
\textbf{ID} & \textbf{Gender} & \textbf{Age} & \textbf{Nation} & \textbf{Game} & \textbf{Students*} & \textbf{Yrs of Exp.} & \textbf{Education} & \textbf{Status} \\
\hline
P1  & Male   & 25  & Portugal    & Rocket League             & 1,000 & 5    & Bachelor's           & Full-time  \\
P2  & Male   & 24  & Bangladesh  & Valorant                  & 200   & 4    & 	Undergrad  & Part-time  \\
P3  & Male   & 24  & Egypt       & League of Legends         & 400   & 5    & Undergrad  & Part-time  \\
P4  & Male   & 26  & Pakistan    & Apex Legends              & 800   & 4    & Master's             & Part-time  \\
P5  & Male   & 27  & Hong Kong   & League of Legends         & 500   & 5    & Bachelor's           & Part-time  \\
P6  & Female & 23  & Mexico      & Chess                     & 100   & 5    & Grad student    & Part-time  \\
P7  & Male   & 21  & Costa Rica  & Overwatch                 & 200   & 5    & Undergrad  & Part-time  \\
P8  & Male   & 27  & Belgium     & FIFA                      & 150   & 4    & Bachelor's           & Full-time  \\
P9  & Male   & 27  & Indonesia   & Dota 2                    & 350   & 5    & Bachelor's           & Full-time  \\
P10 & Male   & 21  & Bulgaria    & Clash Royale              & 150   & $<$1 & Undergrad  & Part-time  \\
P11 & Male   & N/A & Italy       & Teamfight Tactics         & 150   & 3    & N/A                  & Full-time  \\
P12 & Male   & 24  & Pakistan    & Chess                     & 300   & 4    & Grad student    & Part-time  \\
P13 & Male   & 25  & Poland      & CS:GO                     & 500   & 6    & Bachelor's           & Part-time  \\
P14 & Male   & 24  & Turkey      & Marvel Rivals             & 120   & 1    & Bachelor's           & Part-time  \\
P15 & Male   & 33  & Philippines & Hearthstone BG & 150   & 5    & Bachelor's           & Part-time  \\
P16 & Male   & 19  & Pakistan    & Fortnite                  & 90    & 2    & Undergrad  & Part-time  \\
P17 & Male   & 21  & Pakistan    & Minecraft          & 100   & 1    & 	Undergrad  & Part-time  \\
P18 & Male   & 22  & Netherlands & Rainbow Six Siege         & 200   & 7    & Undergrad  & Part-time  \\
P19 & Male   & 24  & Bangladesh  & Valorant                  & 1,000 & 3    & Undergrad  & Part-time  \\
P20 & Male   & 28  & Germany     & Escape from Tarkov        & 150   & 4    & High School          & Part-time  \\
\hline
\multicolumn{9}{l}{\textit{Note.} *Student counts are self-reported approximations. N/A indicates the participant preferred not to disclose.} \\
\end{tabular}
\end{table*}

\subsection{Interview Procedure}
We conducted semi-structured interviews via Zoom and Discord (video format) between November 2025 and February 2026. Each interview lasted approximately 35 minutes and was conducted in English. Since most participants already used English in their regular coaching work, language was not a barrier. All interviews were video-recorded and transcribed using automated transcription tools; the first author then reviewed each transcript against the original recording to correct any errors.
The interview protocol covered four areas. First, we asked participants about their backgrounds, including how they began coaching, the games they coach, and their experience with gig platforms. Second, we explored their coaching practices, including session workflows, the tools they use, and how they manage ongoing relationships with students. Third, we discussed the challenges they encounter in their work, such as difficult student interactions, platform-related frustrations, and their perceptions of coaching as a profession. Fourth, given the growing interest in AI enhanced approaches to game analysis, player development, and even coaching in games ~\cite{bubna2023coaching, xenopoulos2022ggviz,edge2025aipowered}, we asked coaches how they envision AI fitting into their work.

\subsection{Data Analysis}
We analyzed the 20 interview transcripts using reflexive thematic analysis (RTA)~\cite{braun2006using, braun2019reflecting}. The first author familiarized themselves with the data through repeated readings of each transcript and then conducted two rounds of inductive coding, generating an initial set of codes in the first round and refining them in the second. Codes and their supporting extracts were transferred to FigJam~\cite{lucero2015affinity}, where related codes were grouped under candidate theme labels along with associated quotes. The first author led theme development, and all authors then met regularly to interrogate and refine candidate themes, surfacing alternative readings and checking the first author's game-specific interpretations against the broader dataset.

A reflexive concern throughout the analysis was the first author's positionality. The first author has played each of the 17 games represented in the dataset for at least 30 hours, providing the domain familiarity needed to follow participants' game-specific terminology, build rapport during interviews, and conduct meaningful probing. This familiarity also shaped how we read participants' accounts, helping us interpret why certain coaching practices took the forms they did. To keep this familiarity from becoming an implicit baseline, we kept our analytic focus on platform-mediated game coaching practice rather than game-specific coaching: patterns that recurred across structurally different titles were read as features of the freelance coaching ecosystem.
\section{Results}
We present the main themes that emerged from our study.

\subsection{Entering and Sustaining the Game Coaching Gig}
This theme describes how game coaches get started and then stay financially viable on gig platforms.
\subsubsection{How Coaches Enter Freelance Game Coaching}
Coaches described entering game coaching on freelancing platforms as an informal pathway shaped by competitive status, perceived market demand, and platform access. Most began after reaching a high rank and noticing that less skilled players were willing to pay for guidance, then converting that expertise into a gig on digital labor markets. P13 described a typical trajectory: \emph{"I was quite good at the game at the time, reached the highest rank... and just decided I wanna earn some money as a student."} Others [P8, P9, P16] entered coaching when external constraints, e.g., aging out of competition, family caregiving responsibilities, or unreliable infrastructure, made sustained competitive play difficult. Many also began during the COVID-19 period (see Table \ref{tab:demographics}), when online gaming activity increased \cite{ellis2020covid}.

Yet participants emphasized that entering coaching and sustaining it as paid gig work were different challenges. With little formal preparation for teaching or service delivery, coaches learned through experimentation how to package their skills, communicate value, and keep a steady pipeline of students. P8, who transitioned from pro play to coaching amateurs, noted that he \emph{"learned on the go,"} while P17 similarly reflected, \emph{"I've never learned from anyone—just from my experience, I learned."} Early-stage sustainability was often defined by a cold-start period: several coaches [P10, P13, P16, P17] struggled to secure their first paying clients, sometimes waiting months for initial orders. P16 recalled that \emph{"getting your first review is the hardest part"} and offered sessions for as little as \$5 to break through, while P10 created his gig in 2023 but waited nearly two years before orders became consistent.

\subsubsection{How Game Lifecycles Shape Freelance Game Coaching}
Sustaining coaching income also appeared to be constrained by the lifecycle of the specific game being coached. Coaches repeatedly noted that demand is contingent on a title’s popularity and can change abruptly due to trends beyond their control. As P14 explained, \emph{"the game may die, people may move on to another game... so you can't know the future."} P10 similarly experienced an extended period with no orders until the player base of his game grew, after which demand increased. To reduce this volatility, some coaches attempted to migrate across titles, but described only partial portability of expertise. Several [P13, P14, P20] reported that skills transferred more easily within the same genre (e.g., FPS), and P14 expanded from one FPS game to another due to shared mechanics. However, coaches also emphasized that maintaining viability in a new game still required significant reinvestment. P20, who coached an extraction shooter game, acknowledged that even transferring within the same genre, he \emph{"would still need a certain amount of knowledge in the other game, because even though they're the same genre"} the specifics differ substantially. P9 noted a related pressure within a single title: game updates meant coaches must \emph{"always update the resources"} to remain credible, making expertise a continuously depreciating asset rather than a stable credential.

\subsubsection{Adapting to the Metagame Cycle}
Beyond the broader lifecycle of a title, coaches also described expertise as continually reshaped by the metagame cycle of each game. The intensity of this cycle varied markedly across titles in our sample, producing uneven conditions for accumulating coaching knowledge. Chess coaches, working with a game whose rules are fixed and where opening theory evolves without invalidating prior material, drew on stable instructional resources. P6 prepared lessons from published chess books, treating them as durable references, and relied on books organized by skill level to guide students from beginner stages upward. In contrast, coaches of heavily patched, set-based titles described the opposite condition. P15 (Hearthstone Battlegrounds) eventually abandoned a written curriculum, explaining that \emph{"every single patch, you have to change the way you play,"} and P11 (TFT) described the game as transforming wholesale \emph{"every single set and patch."} P9 (Dota 2) reported a similar pressure on a longer timescale, noting that coaching resources from \emph{"one year ago or two years ago"} were no longer reliable.

\subsubsection{Economic Realities and Platform Dependency on Game Coaching}
Coaches’ ability to sustain the gig was shaped by both economic constraints and platform infrastructure. Most participants coached part-time, using it as supplemental income alongside university studies or other jobs; only a few (P1, P8, P9, P11) relied on coaching as a primary livelihood. The global reach of gig platforms also created purchasing power asymmetries that influenced who could make coaching financially worthwhile. Coaches in Egypt, Pakistan, and Indonesia (Global South contexts) described earning locally meaningful income by attracting clients paying rates typical in Western markets. As P3 explained, \emph{"if your English is decent... you could be making \$400 a month, which is a lot in Egypt."} This enabled some non-Western coaches to price competitively while still earning viable wages. At the same time, sustaining coaching work depended heavily on platform-mediated visibility and demand. Nearly all participants relied on the digital labor platform Fiverr to acquire clients, and several [P2, P4, P10, P18] reported difficulty gaining traction on coaching-specific platforms. P4 suggested that platforms focused only on game coaching revenue felt fragile, whereas diversified digital labor platforms, such as Fiverr, offered greater stability. However, this stability came with tradeoffs. Several coaches [P3, P8, P14, P15] described Fiverr's opaque ranking algorithms as a persistent source of stress and uncertainty. P8 captured this instability: \emph{"your visibility goes, skyrockets super high, and then the next day... go extremely low."} Algorithmic penalties could be especially severe; P14 recalled that a single delivery error caused the platform to drastically reduce his visibility, and despite multiple attempts to recover, found it \emph{"too difficult"} to regain his previous standing. These dynamics meant that even established coaches faced income disruption driven by platform infrastructure beyond their control.

Several coaches also developed practices that reduced their dependence on a single platform-mediated revenue stream. P6 and P15 reported moving recurring clients off Fiverr after the first session, citing platform commission fees and slow payment clearance as primary motivations. P8 explicitly framed content creation on external channels as a planned diversification of client acquisition, and P9 expressed similar interest in developing a YouTube or Instagram presence. P19 maintained coaching alongside other businesses rather than relying on it as a primary source of income. These practices varied in form and motivation, but collectively reduced exposure to any single platform or revenue stream.

\subsection{Session Practices and Coaching Approaches}
This theme captures how freelance game coaches rely on self-developed, lightweight practices rather than formal pedagogical frameworks. Despite having no shared training or coordination, coaches converge on a similar session structure shaped by the practical demands of one-on-one remote coaching. At the same time, effective coaching depends strongly on personalization, rapport building, and adaptive communication, which coaches see as their main source of value compared to generic online resources. These practices are supported by minimal and informal tooling, and sustained through ongoing relational work such as follow-up messages, continued support, and community building that extends beyond individual sessions and helps maintain student engagement over time.

\subsubsection{How Freelance Game Coaches Structure Coaching Sessions}
Despite having no shared training or coordination, coaches described a similar session workflow that emerged through practice rather than formal instruction. Sessions generally followed a common progression that supported effective one-on-one remote coaching: initial rapport building, assessment of the student’s background, observation of gameplay either live or recorded, diagnosis of weaknesses, targeted feedback, and a closing summary. For example, P5 described: "\textit{greeting students, having a little bit of chit-chat... ask them to tell me more about themselves... put up one of the replays [replay of how the student plays a game]... ask questions... provide them not the correct answer, but the correct mindset,}" while P8 similarly emphasized breaking the ice before recording gameplay, reviewing footage, and concluding with "\textit{a little summary of the key points.}" While coaches differed in their preferred techniques, with some relying on live gameplay observation [P1, P3, P4, P10, P16] and others on pre-recorded replays [P5, P9, P12, P14, P18], the overall pedagogical flow was strikingly consistent. This convergence suggests that the structure likely arises organically from the needs of individualized remote instruction.

\subsubsection{Personalized and Relationship-Centered Pedagogy}
Coaches consistently framed personalization as central to effective and sustainable coaching. All emphasized adapting instruction to each student’s needs, learning style, and goals. As P1 stated, "\textit{I really try to go around the student's needs because everyone is different,}" and P12 explained that "\textit{chess has to be a personalized thing. Everybody thinks differently.}" Coaches positioned this tailored guidance as the primary value of paid coaching compared to freely available online content. As P19 noted, while "\textit{there's so many stuff on YouTube,}" learners often struggle to identify what applies to them, whereas coaching provides targeted direction. Relationship building was described as equally important to instructional quality. P11 compared coaching effectiveness to medical compliance, stating that "\textit{if they don't like you, even if everything you say is correct, they're not gonna listen to you.}" Coaches described actively adjusting tone, communication style, and pacing based on student reactions. These adaptive pedagogical strategies were developed through experience and ongoing interaction rather than formal training.

\subsubsection{Lightweight Tools and Ongoing Student Management}
Coaches supported their practices with minimal and informal tooling. Most relied on Discord for voice communication and screen sharing, with some supplementing sessions using drawing overlays such as Epic Pen [P7, P8, P18] or game-specific replay systems. Student tracking and note-taking were handled through personal folders [P19], Google Sheets [P9], or memory, reflecting a largely ad hoc approach to student management. Sustaining coaching relationships often extended beyond individual sessions. Several coaches described proactive follow-up practices to encourage repeat engagement. As P7 explained, "\textit{I would hit back all the people that I've been coaching... it usually brings them back.}" Others offered continued messaging support after sessions [P13] or created Discord communities to address recurring questions and maintain contact [P10], although some noted that these approaches did not always scale effectively.

\subsection{Authority and Expectation Conflicts in Freelance Game Coaching}
This theme captures how game coaches navigate ongoing tensions in coach–student interactions that stem from fragile and contested forms of authority. Although students actively seek and pay for coaching, coaches frequently encounter resistance to feedback, particularly from experienced players who equate tenure with expertise. Because game coaching lacks formal credentials, coaches must continuously establish legitimacy through visible markers such as in-game rank, competitive history, and platform ratings. These dynamics place coaches in a precarious position where instructional authority, labor value, and platform reputation must be actively negotiated within each interaction.

\subsubsection{Student Resistance and Fragile Authority in Freelance Game Coaching}
Coaches described student resistance to feedback as a central tension in freelance game coaching, reflecting the fragile and contested nature of coaching authority. Although students voluntarily sought out and paid for coaching, many coaches encountered dismissiveness or contradiction during sessions. P1 recalled students responding with "\textit{Oh, I already know that}," while P2 framed this resistance as identity-level defensiveness: "\textit{when I tell him that what you have been doing your entire life is wrong... maybe as a defensive mechanism, they don't really want to accept it.}" P9 further observed that long-tenured players were especially resistant, often equating years of play with expertise and authority: "\textit{If people say that I've been playing Dota for 20 years, they will think I'm playing Dota more than you, why would I want to listen to you.}"

Coaches linked this resistance to the absence of formal credentials in game coaching, which requires authority to be continuously established rather than assumed. Several coaches [P4, P12, P15, P16] reported that students frequently asked about rank or competitive history before fully accepting feedback. In response, coaches strategically curated visible markers of expertise on their platform profiles—displaying in-game ratings [P12], tournament results via external verification sites [P10], and gameplay footage [P16]. P16 emphasized that showing recorded gameplay was "\textit{really important for the people to see your actual gameplay}," treating video as proof of skill. P4 explained that displaying rare in-game badges gave him "\textit{more authority}" because students needed to believe "\textit{they're doing a session with someone who has actually done it.}" Yet even listed credentials did not settle the question: P4 noted that students "\textit{would actually ask me for proof}" despite his achievements already appearing on his profile, and P15 similarly reported preemptively placing credentials on his page because "\textit{they always ask.}" This pattern reveals that earned authority requires ongoing defense against student skepticism rooted in the shared competitive space coaches and students occupy.

\subsubsection{Devalued Labor and Misaligned Expectations}
Coaches also described how broader devaluation of games as legitimate work intensified authority and expectation conflicts. Several noted that the recreational framing of games shaped perceptions of coaching as less serious or less skilled labor. P18 observed that game itself is "\textit{not really seen as normal}" by broader society, which further undermined the professional legitimacy of coaching, while P14 noted that even full-time coaches on professional teams "\textit{don't get enough respect.}" This devaluation was also reflected in how students valued coaching relative to other game-related spending. As P2 observed, "\textit{they would spend \$100, \$200, \$300, \$500 on skins, but they don't want to get a coach for \$20 or \$30, or even \$10 for an hour.}"

This devaluation directly shaped student expectations in coaching sessions. Some coaches [P8, P9] reported that students expected rapid improvement after a single session and attributed limited progress to coaching failure. As P8 compared, "\textit{if you would take a coaching session with a tennis coach, you wouldn't expect to become Serena Williams in a matter of 2 hours. But for esports, some people had those sort of expectations.}" In a platform-mediated context, these misaligned expectations carried material consequences, as negative reviews or refund demands directly affected coaches’ visibility, reputation, and ability to sustain work on freelancing platforms.

\subsection{Perspectives on AI in Game Coaching}
This theme describes how freelance game coaches make sense of AI in their work by drawing a clear boundary between what they see as non-negotiably human and what they are willing to delegate to automation.

\subsubsection{Encounters with AI Coaching Tools in Practice}  
Although our interview protocol asked broadly about the tools coaches used in their work, beyond general communication infrastructure such as Discord and Zoom and game-specific utilities such as Epic Pen and built-in replay viewers, no other computational tools surfaced as recurrent features of practice. AI was the exception: while only a few coaches reported direct hands-on encounters with AI-based tools, nearly all participants had formed views on AI's role in coaching, drawing on adjacent observations and reflections about their own work. Coaches' perspectives on AI emerged not from speculation about hypothetical systems but from direct encounters and adjacent observations. P16 described Osirion\footnote{Osirion (\url{https://osirion.gg}) is a commercial AI-powered improvement platform for Fortnite that provides custom game plans, statistics, and AI coaching feedback.}, an AI-powered website for Fortnite that generates practice routines based on player data, calling it \emph{``the best thing for Fortnite''} while noting its current limitations: \emph{``right now it's bad, because AI is still in the process of developing.''} P12 reflected on chess engines such as Stockfish as a potential coaching aid given the fixed, well-analyzed nature of chess. P9 (Dota 2) observed content creators consulting general-purpose chatbots such as ChatGPT for in-game advice, including item builds, and assessed the limits of this use case directly: in Dota, item choices must continuously adapt to opponents' purchases and game state, decisions that unfold faster than any input-output exchange with a chatbot can accommodate. P9 noted that even the content creator using ChatGPT did not follow its recommendations wholesale, selecting only one or two items before reverting to in-game judgment. These encounters varied in form, spanning game-specific training websites, long-standing game engines, and general-purpose chatbots, but shared a common feature: AI-based tools had already entered the surroundings of game learning infrastructure. The two subsections that follow describe how coaches assessed this landscape, distinguishing between roles they resisted and roles they welcomed.

\subsubsection{Skepticism Toward Automated Coaching}
Most coaches were skeptical that AI could replace human coaching, citing frequent game updates and the complexity of real-time gameplay. Coaches described how developer patches regularly invalidate existing strategies, requiring constant adaptation. As P9 noted, "\textit{you need to always update the resources because your data is like not really updated.}" P7 similarly observed that "\textit{the meta changes every time... because people adapt, people get better.}" Beyond handling game updates, coaches emphasized how difficult it would be for an AI to analyze the many shifting variables in competitive gameplay. Several [P4, P8, P14] argued that an AI coach would likely produce generic tips rather than the individualized diagnosis students pay for. P14 anticipated that AI "\textit{would give too many general tips, basically, not specific to the person,}" while P4 captured the deeper limitation: "\textit{You cannot teach AI what is not even known to the internet.}" However, chess coaches provided a notable contrast. Likely because chess rules remain fixed and algorithmic analysis is well established \cite{silver2018general}, participants, like P12, acknowledged that if an AI engine "\textit{can also communicate, then yeah, it has exceeded}" human coaching, and P6 conceded AI "\textit{could teach chess.}" In other games, participants [P9, P16, P18] suggested that AI coaching might be most useful for beginners by providing foundational support, such as explaining mechanics or terminology, rather than delivering advanced and personalized coaching. Even so, they emphasized that AI could not replicate the interpersonal warmth or motivational support that human coaches provide.

\subsubsection{Desired Role of AI as Coaching Assistant}
While coaches were skeptical of AI taking on a coaching role and actively participating in their coaching sessions, many were receptive to it as a behind-the-scenes workflow assistant. Before a coaching session, coaches expressed interest in AI that could handle scheduling [P7], answer frequently asked questions [P7, P18], and gather background information from students [P20]. Yet this openness came with clear boundaries shaped by negative past experiences. P12 disabled an AI response tool after it replied to clients without notifying him, and P17 noted that \textit{``people just didn't trust my coaching after seeing that AI talking to them instead of me,''} making clear that any client-facing automation needed to preserve the coach's personal presence and keep him informed. After coaching sessions, coaches expressed similar interest in AI support for administrative work. Generating session summaries, organizing notes, and tracking lesson plans were seen as necessary but time-consuming tasks, and coaches wanted AI to help shoulder this burden [P3, P12, P14, P19]. P12 captured this appeal clearly, imagining an AI that would keep \textit{``updating that person's file''} after every lesson, reducing documentation work without interfering with the coach--student relationship. Across both phases (pre and post coaching phases), coaches consistently drew the same line: AI was welcome when it reduced administrative overhead, but not when it risked displacing the direct interaction through which they built trust and authority with students. This likely reflects a broader tension in freelance game coaching, where practitioners must repeatedly establish credibility in each new session while navigating volatile platform and game ecosystems. In that context, AI earns acceptance when it supports existing work from the background, and meets resistance the moment it steps into the space where the coach--student relationship is formed.

\section{Discussion}
Studying freelance game coaching clarifies what is at stake as Games HCI builds computational tools to support game learning \cite{lee2025crafting}, an effort that prior work shows is not straightforward \cite{kleinman2026really}. Before designing systems that augment or replace human coaches, Games HCI needs a clearer account of how those coaches actually work, and how they earn the position to advise players on their gameplay. We focus on experienced coaches because these coaches have likely already developed more stable working methods and formed considered views on what helps players improve. Throughout our interview study we uncovered that freelance game coaching is not simply a matter of transmitting gameplay knowledge, but a configuration of work in which platform-mediated labor, the lifecycle of specific game titles, and the relational dynamics of skill instruction are tightly coupled.

Across the 20 coaches in our study, this coupling produced patterns that neither platform labor research \cite{kalleberg2009precarious, wood2019good} nor existing accounts of game coaching \cite{watson2025introducing} fully anticipate. Coaches experienced precarity across two independent axes: the instability of gig platform visibility systems and the volatility of the games themselves, including shifting metas \cite{kokkinakis2021metagaming}, and player populations \cite{demediuk2018player}. At the same time, their instructional authority could not rest on formal credentials. Instead, it had to be continuously demonstrated within the same competitive systems in which their students also participated. In what follows, we discuss these two themes that emerge from our findings, dual precarity and earned authority, and conclude with implications for the design and integration of coaching tools that complement, rather than displace, the situated expertise and relational labor at the center of this work.

\subsection{Dual Precarity}
Prior research characterizes platform labor as precarious \cite{kalleberg2009precarious, wood2019good}. Our findings align with this work, but extend ongoing CHI PLAY discussions about the relationship between games, labor, and live-service infrastructures by revealing an additional structural vulnerability. Unlike many forms of gig work, freelance game coaching is simultaneously embedded within a digital labor marketplace and within the evolving ecosystem of a specific game title, creating what we term \emph{``dual precarity''}. Our interviews revealed how coaches must navigate both the familiar instabilities of platform-mediated labor and the lifecycle volatility of live-service games, where updates and declining player populations can rapidly reshape demand for coaching \cite{demediuk2018player}.

Although some coaching competencies might transfer across games within the same genre, our interviews also highlighted how coaches still must repeatedly reinvest time into learning title-specific mechanics, strategies, and community norms. This can limit the portability of their expertise and ties their livelihood to the ongoing health of particular game ecosystems. Unlike coaches employed by professional esports organizations \cite{watson2025introducing, lee2025crafting}, who may receive salaries that buffer short-term fluctuations in a game's popularity, freelance coaches experience these shifts immediately: as a game's player base contracts, so does their pool of potential students. In this sense, the sustainability of game coaching becomes tightly coupled to the maintenance practices and commercial decisions surrounding live-service games themselves.

Importantly, these two forms of precarity do not operate independently. Stability on one axis cannot compensate for instability on the other: a thriving game ecosystem cannot protect coaches from algorithmic deranking on gig platforms [P3, P8, P14], and, conversely, a strong gig platform reputation cannot sustain work when a game's player base collapses [P10, P14, P18]. Freelance coaches must therefore continually manage both forms of instability simultaneously, without the institutional protections available in more formalized esports settings. This layered dependency highlights how coaching labor in games is shaped not only by player skill and pedagogy, but also by the broader socio-technical infrastructures through which games are distributed, updated, and monetized.

\subsubsection{Adapting to the Metagame Cycle}
Prior esports literature defines the metagame as the dominant set of in-game strategies that prevails at a given point in time \cite{kokkinakis2021metagaming}. Because live-service games are continually patched, this dominant strategy shifts repeatedly, producing a recurring cycle that coaches must track. Our findings show that this cycle has direct consequences for coaching labor. As games are updated, patched, and rebalanced, coaches must continually revise the knowledge they use to diagnose player weaknesses and provide guidance. Their expertise is therefore not simply accumulated over time; it must be repeatedly maintained, updated, and sometimes relearned. This broadens the scope of Games HCI inquiry by showing that game updates and lifecycle shifts affect not only players, but also the workers who support player learning. Coaches’ knowledge can expire when developers change the game, and the speed of that expiration varies by title. Coaches therefore operate within a layered dependency structure in which both freelance platform algorithms and game publisher decisions can undermine their professional standing. Design support for these coaches should not assume that coaching materials, curricula, or expertise remain stable. For many titles, the more urgent need might be lightweight infrastructure tracking gameplay updates, revising training materials, or tools that could help them adapt coaching strategies to shifting metas.

\subsubsection{Coach Strategies for Managing Dual Precarity}
Although coaches in our study did not always explicitly describe their practices as forms of risk management, many of their strategies functioned to reduce exposure to dual precarity. These practices show how players-turned-coaches adapt to the unstable socio-technical infrastructures surrounding live-service games and platform-mediated gaming labor. To reduce dependence on platform-level instability, several coaches described developing practices that weakened reliance on Fiverr as a single source of visibility or income. They moved recurring clients off-platform, planned content creation on external channels to support future client acquisition, or treated coaching as one income stream among several rather than as a sole livelihood. Coaches also shared how they worked to extend the value of player relationships beyond one-time coaching sessions. Practices such as initiating follow-up contact, offering open-ended post-session support, and maintaining student communities created forms of continuity and retention that were not entirely dependent on platform algorithms.

Against title-level volatility, coaches described how expertise required continual reinvestment to remain current, with patches and content updates regularly invalidating prior knowledge [P9, P11, P15]. Some coaches [P13, P14, P20] described limited portability across structurally adjacent games within the same genre, but even this transfer required substantial title-specific relearning. These findings extend CHI PLAY discussions about player communities and live-service game ecosystems \cite{larsen2024community, dubois2022games}, by showing how developers' ongoing production decisions reshape not only player experience but also the livelihoods of freelance coaches who teach within these games.

Importantly, these responses do not eliminate the structural pressures of dual precarity, but instead suggest that freelance coaches continuously negotiate instability across platform, relational, and gameplay expertise layers. This has implications for the design of future coaching technologies in Games HCI. Rather than positioning coaches as passive recipients of technological support, or attempting to replace the adaptive social practices they already rely on, such systems should aim to strengthen the coordination, community-building, and cross-title adaptation strategies that coaches have developed within evolving gaming ecosystems.

\subsection{Relational Gameplay Diagnosis in Freelance Coaching}
All coaches in our study located their expertise in individualized diagnosis: reading a learner’s gameplay, identifying underlying problems, and calibrating guidance to that player’s needs, personality, and goals. This frames game coaching not simply as knowledge transfer, but as relational gameplay support shaped by the social and economic conditions of play. It also diverges from salaried esports coaching, where Watson et al. find that team-employed coaches often rely on coach-centric, solution-driven instruction due to competitive pressure and limited formal coach education \cite{watson2025introducing}. Freelance coaches operated under different relational conditions. Their students were paying clients who could disengage, request refunds, or leave reviews affecting platform visibility. As a result, feedback had to be carefully calibrated: direct criticism risked losing students [P8, P18, P19], so coaches softened corrections, used questions rather than commands [P5, P7], and built rapport before critique [P4, P8, P10]. Tone and communication style became central to coaching, adapted to students’ expectations and emotional responses [P4, P12]. In freelance game coaching, diagnosis cannot be separated from trust and legitimacy. Unlike salaried coaches, who can rely on institutional hierarchy, freelance coaches teach students who are simultaneously learners, customers, and evaluators. Effective coaching emerges from the interplay between gameplay expertise and sustained relational engagement within platform-mediated interactions.

Relational engagement in freelance coaching may also be shaped by gender. The one woman coach in our sample [P6] described interactions that departed from the skill-centered dynamics other coaches reported, where some prospective clients approached her on the basis of her gender rather than to be coached, while in child-oriented coaching parents often preferred a woman coach for their children. These accounts come from a single participant and cannot be characterized systematically here, but they suggest that the relational conditions through which coaching unfolds are not gender-neutral, an aspect future work could examine directly.

\subsection{Earned Authority and the Negotiation of Coaching Expertise in Competitive Gaming}
Beyond the structural precarity outlined above, coaches also faced a distinct relational challenge: our interviews highlighted how authority in freelance game coaching is not granted through institutional credentials, but must be continually established through visible performance within the game itself. Although students voluntarily sought out and paid for coaching, coaches repeatedly reported that students asked about a coach's rank, in-game achievements, or prior competitive experience before fully accepting feedback [P4, P12, P15, P16]. Instructional legitimacy therefore emerged from demonstrated participation and success within a shared competitive ecosystem rather than from externally certified expertise. This differs from professions where coaching authority is institutionally stabilized \cite{watson2025introducing, lee2025crafting}. In freelance game coaching, coaches and students inhabit the same competitive activity, sharing experiences of ranking systems, matchmaking, and competitive progression. As P8 observed, a tennis student would not expect to rival a coach after a single lesson, but in competitive games, the shared competitive space leads students to position their own gameplay experience as grounds for evaluating or contesting coaching advice. This dynamic reflects a broader feature of competitive gaming cultures in which expertise remains highly visible, continuously measured, and open to challenge through in-game performance itself.

To navigate this environment, participants in our study described how they constructed authority through recognizable signals embedded within gaming ecosystems: displaying their game ranks, in-game badges, tournament histories, gameplay footage, or affiliations with competitive teams on their platform profiles. Yet, our participants also described how these signals rarely settled legitimacy permanently. Students often requested additional proof of skill beyond profile information, and students with extensive gameplay histories were especially likely to position their own experience as a competing basis for judgment [P1, P2, P3, P5, P8, P9, P18, P19]. Earned authority therefore remained relational and ongoing, requiring coaches to repeatedly demonstrate expertise through interaction, communication, and continued competitive relevance. Competitive games create shared spaces where students can directly compare themselves to instructors through rankings, mechanics, and gameplay knowledge. The result is a form of instructional labor in which coaching expertise is persistently contestable and socially negotiated.

We also observed that alternative pathways to stable authority did exist, but only in limited forms. A small number of coaches in our study described using professional-level accomplishments to reduce the need to repeatedly establish legitimacy with students. These included affiliations with esports organizations, internationally recognized rankings, or high-level competitive achievements [P10, P12, P16]. However, such credentials were rare. Most coaches in our study operated instead as highly ranked but non-professional players. Formal coaching certification was not visible at all. Although national esports federations have developed coaching licenses in some contexts,\footnote{For example, the ESBD's E-Sport Trainer in Germany~\cite{esbd} and KeSPA's e-sports coaching license in South Korea~\cite{kespa}.} none of the coaches in our study referenced these credentials, and no Fiverr profiles from the participants displayed them. This aligns with Watson et al.~\cite{watson2025introducing}, who observe that even professional esports coaches often rely on playing experience and self-taught methods rather than formal pedagogical training or coaching licenses. This relational fragility compounds the structural precarity described earlier. Freelance coaches must simultaneously manage gig platform visibility, shifting game ecosystems, and ongoing challenges to their legitimacy from students embedded in the same competitive environments. For Games HCI, these findings reframe coaching in games beyond pedagogy or skill transfer, foregrounding the socio-technical negotiation of expertise and trust within competitive game play cultures.

\subsection{Design Implications}
Our findings highlight a design opportunity for coaching tools in Games HCI. Rather than treating freelance coaches as external to the game experience, future systems could use AI or other computational tools to support the coordination and analytic work that makes coaching possible. Coaches in our study were receptive to AI when it reduced administrative overhead while preserving the coach--student relationship. They welcomed AI support for scheduling, session summarization, note organization, progress tracking, and follow-up communication management, tasks they currently handled manually through spreadsheets, folders, Discord messages, or memory. AI and other computational tools could also support coaching-specific analytics by surfacing replay moments tied to a student’s recurring patterns~\cite{kleinman2022time, xenopoulos2022ggviz}, tracking longitudinal progress across patches, organizing replay annotations, flagging lesson plans that reference outdated strategies~\cite{kokkinakis2021metagaming}, or supporting credential display and competitive benchmarking. In doing so, systems could help coaches navigate aspects of the dual precarity by reducing administrative burden, preserving coaching histories, and helping coaches stay current as games evolve. At the same time, our findings caution against framing AI as a replacement for human coaches. Most coaches were skeptical that AI could handle the complexity and variability of real-time gameplay [P4, P7, P8, P9, P14], and those who experimented with client-facing automation reported negative student reactions once AI involvement became visible [P12, P17]. Prior work on tutoring similarly shows that effective instruction depends not only on analytical correctness, but also on relational and conversational support, including scaffolding reasoning, sustaining motivation, and building trust~\cite{vanlehn2011relative, thomas2024improving}.

Together, our findings point toward hybrid game coaching rather than full AI replacement. AI may be most effective when augmenting the scaffolding around expert gameplay diagnosis rather than replacing the diagnostic interaction itself~\cite{wang2024feature, xenopoulos2022ggviz}. In such systems, AI could support benchmarking, longitudinal pattern detection, adaptive lesson planning, and administrative coordination, while human coaches retain the relational work of timing criticism, building rapport, responding to learners’ emotional states, and interpreting gameplay patterns within their broader context~\cite{thomas2024improving}.

\subsubsection{Designing Coaching Systems Around Earned Authority}
Prior work shows that esports tools are accepted as augmentation but resisted as replacement \cite{kleinman2026really}. Our findings suggest this resistance is tied to how authority is established in competitive gaming cultures. Future coaching tools may need to make competence legible through visible benchmarking or match histories within a competitive ecosystem. Chess illustrates this dynamic: because AI systems have demonstrated superhuman competitive performance \cite{silver2018general}, chess engines are widely accepted as legitimate analytical tools by players and coaches [P6, P12]. Our findings also suggest that tools could help freelance coaches construct and maintain authority. Coaches currently rely on fragmented signals such as ranks, badges, tournament histories, gameplay clips, and reviews spread across platforms. Future systems could aggregate these signals into verified coaching portfolios, track competitive achievements, surface coaching outcomes, visualize student improvement trajectories, or show adaptive lesson histories tied to changing metas. Such tools could reduce the burden coaches face when repeatedly proving legitimacy to new students.

\subsubsection{Designing Patch-Aware Coaching Tools}
Our findings also suggest that Games HCI systems for competitive learning should treat coaching knowledge as dynamic rather than static. The expertise coaches produce, maintain, and transmit cannot be fully captured in a fixed guide or encoded once into a tool. Instead, coaching expertise is an adaptive practice shaped by patches, shifting metas, and changing player strategies. This creates an opportunity for coaching tools that help coaches maintain and revise their knowledge as games evolve. Future tools could help coaches track when game updates alter relevant strategies, flag coaching materials that may have become outdated, summarize patch changes in relation to a coach's existing lesson plans, and flag which replay may need revision. Computational tools could also help preserve how coaching advice changes across game versions, allowing coaches and students to see how learning goals, tactics, and player-development trajectories shift over time. Designing for competitive learning in live-service games therefore requires systems that provide coach-facing infrastructures for maintaining expertise as the game itself changes beneath them.

\subsection{Limitations}
This study has several limitations that bound the scope of our findings. Our participants operated primarily through Fiverr, a platform whose user base skews heavily toward English-speaking markets. Accordingly, the majority of coaches served students from the United States and Europe, and all coaching was conducted in English. The practices and challenges of coaches operating in non-English-speaking contexts, or outside established platforms, remain unexamined. Our recruitment criteria ensured demonstrable platform experience but also privileged coaches who had already achieved a degree of visibility. Coaches operating below this threshold, including those still struggling to secure initial orders and those who migrated to alternative channels, fall outside our sample. The distinct barriers that characterize this earlier stage are therefore not represented in our findings. This gap points to a direction for future work. A coach's competitive skill does not automatically translate into the ability to attract students, communicate effectively, or sustain income on a platform. Future research could examine this gap directly: what supports help a highly skilled player convert expertise into a viable coaching practice, and where do most aspiring coaches stall before reaching that point. Investigating the tools, platform mechanics, and forms of guidance that lower the barrier to entry at this earliest stage could help make coaching a more accessible form of work for skilled players who lack an established foothold.

Our sample was also heavily skewed by gender. Existing scholarship documents how women in esports and competitive gaming face gendered barriers to recognition as high-performing players, including harassment, skepticism toward their competence, and broader masculine-coded community norms~\cite{siutila2019pure, fox2014sexism}. The woman coach in our sample [P6] reported encounters consistent with this literature. Her experience suggests that the relational conditions of coaching are not gender-neutral, but a single participant cannot support systematic claims about how gender shapes this work. Our recruitment did not stratify by gender, and coach gender is often not visible on Fiverr listings, which left the resulting sample heavily skewed. Future work should examine the experiences of women coaches directly, recruiting purposively to understand how gender shapes coaching practice in this setting.
\section{Conclusion}
This paper presented a qualitative interview study of 20 experienced freelance game coaches working across 17 competitive titles. Despite operating independently, coaches converged on a pedagogical practice centered on diagnosing situated gameplay and translating competitive expertise into adaptive feedback. This infrastructure is fragile. Coaches must continually earn authority within the same competitive spaces as their students while navigating dual precarity across platform algorithms and game lifecycles. These conditions shape where AI support can meaningfully enter: coaches welcomed AI for behind-the-scenes administrative work, but resisted it in diagnostic interactions where authority, trust, and relational expertise are built. As AI-supported player-development tools become more common, games HCI research should examine how such systems can support freelance coaching ecosystems without displacing the relational and diagnostic work that makes coaching valuable.

{\bf Acknowledgments.} Special thanks to the freelance coaches who participated in the study and the anonymous reviewers. This work was partially supported by NSF grants 2339443 and 2403252.

\bibliographystyle{ACM-Reference-Format}
\bibliography{refs}

\end{document}